# Diffusion of Mn through GaAs barrier

J. Adell[1], I. Ulfat[1,2], L. Ilver[1], J. Sadowski[3,4], and J. Kanski[1]

[1] Department of Applied Physics, Chalmers University of Technology, SE-412 96 Gothenburg, Sweden

[2] Department of Physics, University of Karachi, Karachi-75270, Pakistan

[3] Institute of Physics, Polish Academy of Sciences, PL-02-668 Warsaw, Poland

[4] MAX-lab, Lund University, SE-221 00 Lund, Sweden

ABSTRACT

Thermally stimulated diffusion of Mn across the (Ga,Mn)As/GaAs interface has been studied by X-ray photoemission. Three $Ga_{0.95}Mn_{0.05}As$ were capped with GaAs of different thickness 4, 6 and 8ML, and Mn diffusing through the GaAs layers was trapped on the surface by means of amorphous As covering the surface. It was found that the out diffusion is strongly reduced for the 6 ML GaAs film, and for the 8 ML film no Mn could be detected. Our results are interpreted as an effect of an electrostatic barrier formed by the GaAs layer.

**Introduction**

It is well known that the ferromagnetic state of (Ga,Mn)As is significantly stabilized by low-temperature (LT) post-growth annealing [1-4]. The increase of the Curie temperature from typically 50-80 K for as-grown layers [5, 6] to the present record high value of around 191 K [4] is ascribed to removal of Mn interstitials ($Mn_I$). Since the magnetic properties are not at all affected if the LT annealing is made without any reactive surface agent, it appears clear that diffusion of the interstitial Mn into the underlying GaAs is efficiently hindered. The inability of Mn to diffuse through GaAs has also been observed in experiments on (Ga,Mn)As capped with thin GaAs films, in which case the effect of post-growth annealing was eliminated for 10 ML thick capping [7]. On the other hand, cross-sectional STM images from (Ga,Mn)As/GaAs multilayer structures suggest that the Mn concentration in 60 ML thick GaAs spacers may be as high as 10-20% of the total Mn concentration in the (Ga,Mn)As after annealing treatment [8], and a depth profiling study of a Mn/GaAs interface indicated massive diffusion of Mn into GaAs [9]. These apparently contradictive pictures require further studies of the (Ga,Mn)As/Ga interface.

The aim of the present study is to establish the diffusion range of Mn in GaAs overlayers on (Ga,Mn)As during post-growth annealing. To detect the out-diffused Mn atoms we have applied a "fly paper" method, in which diffusing Mn atoms that reach the surface of the GaAs overlayer are trapped by bonding to amorphous As that covers the surface. We have demonstrated earlier that this method is orders of magnitude more efficient than in-air annealing for removing Mn interstitials [10] and that the process is saturated upon accumulation of about one monolayer (ML) MnAs [11]. Apart from being a very efficient and direct method for detecting the out-diffused Mn, an advantage of this procedure as compared with depth profiling is that it avoids potential complications due to knock-in effects.

**Experiments**

The photoemission experiments were carried out at the Swedish synchrotron radiation laboratory MAX-lab, beamline D1011 [12]. The samples were prepared by molecular beam epitaxy (MBE) in a local growth system. A small ion-pumped transport vessel was used to bring the samples in ultrahigh vacuum (UHV) to the analysis station. This is essential because (Ga,Mn)As is intrinsically metastable, so its surface cannot be prepared in a clean and well-characterized state by sputtering and annealing. Each sample was transferred three times, first for examination of the as-grown GaAs-covered surface then back to the MBE chamber for annealing under As, and finally for studies of the annealed sample. The (Ga,Mn)As layers were grown on 1x1 cm$^2$ pieces of epiready n-type GaAs(100), glued with indium on transferable Mo holders. In the MBE system the Mo-holders were heated radiatively from the backside by a tungsten filament and the sample temperature was measured with an IR pyrometer. The Ga and Mn beam fluxes were calibrated by means of RHEED oscillations. After conventional oxide desorption and growth of a GaAs buffer at 580 °C, the substrate temperature was reduced to 230 °C and the As$_2$/Ga pressure ratio was set to ~1.5 for the growth of (Ga,Mn)As. Four Ga$_{0.95}$Mn$_{0.05}$As samples were grown and investigated, each 500 Å thick. Three of these samples were capped with LT-GaAs, 4, 6 and 8 ML thick, respectively, and one was left uncapped. The LT-GaAs was grown by closing the Mn shutter at the end of the (Ga,Mn)As growth. Each post-growth annealing was carried out at 210 °C and lasted for 3 hours. After the annealing periods the remaining As was desorbed by raising the substrate temperature for a few minutes to around 250 °C until a clear RHEED pattern was restored.

The photoemission spectra were recorded in normal emission using 1000 eV photons and operating the electron energy analyzer (Scienta SES200) at a constant pass energy of 100 eV. The overall energy resolution was 0.4 eV. All spectra were recorded with the sample at room temperature.

**Results and discussion**

Survey spectra from the different samples recorded prior to annealing are shown in Fig. 1. The spectra were normalized at the Ga3d peak. Despite precautions with keeping the sample in UHV throughout the experiments, it is clear that some surface oxidation has taken place, confirming the strong reactivity of surfaces containing Mn. The oxidation is most likely caused by short exposures to pressure in the high $10^{-8}$ torr range occurring at operation of gate valves during the transfer. For the much less reactive GaAs surface no traces of oxidation can be discerned. As expected, the spectra show gradually diminishing Mn2p emission with increasing GaAs capping thickness. From the attenuation of the Mn2p emission (inset Fig. 1) we can extract a mean free path in GaAs at 365 eV kinetic energy of 1.2 nm, in good agreement with the universal curve of electron mean free path [13].

As already mentioned, annealing of (Ga,Mn)As without any reactive surface agent does not improve the magnetic properties, and photoemission spectra from such annealed samples are practically unaffected by the annealing what concerns the Mn2p intensity. In contrast, annealing of As-capped (Ga,Mn)As can result in enhanced Mn2p emission by up to 4-5 times, depending on the thickness and preparation conditions of the (Ga,Mn)As layer. The enhancement is due to diffusion and bonding of interstitial Mn at the (Ga,Mn)As/As interface, and the maximum enhancement corresponds well to accumulation of a full monolayer of MnAs [11]. In Fig. 2, top curves, we see that in the present case the Mn2p emission from the "No GaAs" sample is increased after annealing, though the increase is not as large. Adopting the same model as in ref. 11 we estimate that the Mn2p intensity increase after annealing corresponds to about 25% of a monolayer MnAs on top of the annealed surface, clearly less than the expected full monolayer. Since the out-diffusion process has been demonstrated to be efficient [11], the relatively low enhancement must be explained by a relatively low content of Mn interstitials in the (Ga,Mn)As film. Nevertheless, there is a clear enhancement, which reflects Mn accumulation on the surface. It is worth stressing that our calibration of the Mn content is based on RHEED oscillations, and is therefore only sensitive to Mn in substitutional sites. The relative amount of Mn in interstitial sites depends on the detailed growth parameters, and is probably the main reason for uncontrolled variations in the magnetic properties of as-grown layers.

The limited control of interstitial Mn is also manifested in the spectra from the sample capped with 4 ML GaAs: prior to annealing the Mn intensity is attenuated by the capping, in agreement with the 1.2 nm mean free path, but after annealing the intensity is actually somewhat larger than for the uncapped sample. With the same assumptions as above, the Mn2p intensity after annealing corresponds to about 50% of a full MnAs layer. The concentration of interstitial Mn is thus somewhat higher in this particular (Ga,Mn)As sample. Due to these variations it is not possible to draw quantitative conclusions regarding the diffusion process. Nevertheless the data show, beyond any doubt, that Mn interstitials can diffuse through 4 ML GaAs.

For samples with thicker GaAs overlayers the situation is clearly different. With 6ML GaAs capping layer the emission from Mn2p before annealing is initially further reduced. After annealing only a slight enhancement is observed. Assuming that the increase is entirely due to surface accumulation, the enhancement corresponds to around 0.05 ML of MnAs on top of the GaAs surface, showing that just a small fraction of the Mn interstitials diffuse through the GaAs layer during the annealing process. Finally, with 8 ML GaAs capping we see that the Mn2p emission is not at all affected by the post-growth annealing. This is seen not only at the peak intensities but also at the low-energy background levels. The complete absence of Mn accumulation shows not only that Mn does not reach the surface, but indicates also that whatever Mn has diffused to the near-surface region of the GaAs layer has moved back during the cooling to room temperature.

We can conclude that there is a thickness-dependent mechanism that very efficiently stops the diffusing Mn interstitials from reaching the surface when the GaAs overlayer thickness approaches 10 atomic layers, i.e. around 2.5 nm. This result is in very good agreement with the previous notion that the beneficial post-growth annealing effects on the ferromagnetic transition temperature are suppressed by 10 ML thick GaAs capping [9]. The rapid decrease of the diffusion efficiency can be directly ascribed to the development of an electrostatic barrier. The gradual change of the surface potential, determined by Fermi level pinning measurements [14], is shown in Fig. 3. The observed pinning position at 0.5 eV above VBM indicates that the pinning is caused by As antisite defects [15], and the origin of the short-range band-bending is the high concentration of such defects in LT GaAs [16, 17]. In the interface region the defects become ionized due to diffusion of holes from the (Ga,Mn)As. The consistency of our observations can be checked using standard expressions for interface electrostatics [18]. With a concentration of As antisites in the range of 0.5 - 1 % [16, 17] and around 0.5 eV potential difference, one obtains a "depletion layer" width of around 2 nm, in good agreement with our observations (Fig. 3). It can be concluded that the diffusing $Mn^{2+}$ interstitials are prevented from reaching the surface by the positive potential barrier that develops with increasing GaAs thickness and that the transport of Mn ions towards the surface is a net effect of thermally stimulated diffusion and drift in the opposite direction due to the band bending. The dynamics of this process is rather complex due to the presumably short mean free path for the diffusing Mn ions, which means that the ions are partially thermalized before moving away from a site. Nevertheless the model can account qualitatively for the present results. Apart from the reduced rate of diffusion with increasing GaAs thickness, we note that the absence of in-diffused Mn in the 8 ML thick GaAs layer implies that the diffusion profile is driven back as the temperature is lowered. This means that the interstitial Mn is quite mobile also at significantly lower temperature than that applied during the annealing process.

We note finally that similar band bending situations are expected at buried interfaces between (Ga,Mn)As and n-type or undoped GaAs. This would contradict the results in ref. 8, where massive Mn diffusion into GaAs was concluded from cross section STM on superlattice structures. To explain this discrepancy, we can speculate that the observed diffusion really occurred at the cleaved surface and not through the whole layer. Likewise, we must conclude that the diffusion reported between Mn and GaAs [9], where a similar band bending situation should occur, was due to ion beam related knock-in effects rather than thermal diffusion.

**Summary**

We have shown that diffusion of Mn across the (Ga,Mn)As/GaAs interface is limited to a range of less than 10 monolayers. The observations are explained as a result of an electrostatic potential developing due to surface Fermi level pinning.

**Acknowledgement**

The present work is part of a project supported by the Swedish Research Council (VR). One of the authors (J.S.) acknowledges the financial support from the Ministry of Science and Higher Education (Poland) through grant N N202 126035.

[1] K. C. Ku, S. J. Potashnik, R. F. Wang, S. H. Chun, P. Schiffer, N. Samarth, M. J. Seong, A. Mascarenhas, E. Johnston-Halperin, R. C. Myers, A. C. Gossard and D. D. Awschalom, Appl. Phys. Lett. **82**, 2302 (2003).

[2] D. Chiba, K. Takamura, F. Matsukura and H. Ohno, Appl. Phys. Lett. **82**, 3020 (2003).

[3] B.J. Kirby, J.A. Borchers, J.J. Rhyne, S.G.E. te Velthuis, A. Hoffmann, K.V. O'Donovan, T. Wojtowicz, X. Liu, W.L. Lim and J.K. Furdyna, Phys. Rev. B **69**, 081307 (2004).

[4] L. Chen, S. Yan, P. F. Xu, J. Lu, W. Z. Wang, J. J. Deng, X. Qian, Y. Ji, and J. H. Zhao, Appl. Phys. Lett. **95**, 182505 (2009).

[5] H. Ohno, A. Shen, F. Matsukura, A. Oiwa, A. Endo, S. Katsumoto and Y. Iye, Appl. Phys. Lett, **69**, 363 (1996).

[6] J. Sadowski, J.Z. Domagala, J. Bak-Misiuk, S. Kolesnik, M. Sawicki, K. Swiatek, J. Kanski, L. Ilver and V. Ström, J. Vac. Sci. Technol. B **18**, 1967 (2000).

[7] M. B. Stone, K. C. Ku, S. J. Potashnik, B. L. Sheu, N. Samarth and P. Schiffer, Appl. Phys. Lett. **83**, 4568 (2003).

[8] A. Mikkelsen and E. Lundgren, Progress in Surf. Sci. **103**, 1–25 (2005).

[9] Y. Osafune, G. S. Song, J. I. Hwang, Y. Ishida, M. Kobayashi, K. Ebata, Y. Ooki, A. Fujimori, J. Okabayashi, K. Kanai, K. Kubo and M. Oshima, J. Appl. Phys. **103**, 103717 (2008).

[10] M. Adell, L. Ilver, J. Kanski, V. Stanciu, P. Svedlindh, J. Sadowski, J.Z. Domagala, F. Terki C. Hernandez and S. Charar, Appl. Phys. Lett. **86**, 112501 (2005).

[11] M. Adell, J. Adell, L. Ilver, J. Kanski, J. Sadowski and J.Z. Domagala, Phys. Rev. B **75,** 054415 (2007).

[12] http://www.maxlab.lu.se/beamline/max-ii/d1011/d1011.html

[13] H. Gant and W. Mönch, Surface Science **105**. 217 (1981).

[14] M. Adell, J. Adell, L. Ilver, J. Kanski and J. Sadowski, Appl. Phys. Lett. **89**, 172509 (2006).

[15] Z.R.M. Feenstra, J.M. Woodall, and G.D. Pettit, Phys. Rev. Lett. **71**, 1176 (1993)

[16] H. Åsklund, L. Ilver, J. Kanski, J. Sadowski, and M. Karlsteen, Phys. Rev. B **65**, 115335

(2002).

[17] X. Liu, A. Prasad, J. Nishio, E.R. Weber, Z. Liliental-Weber, and W. Walukiewicz, Appl. Phys. Lett. **67**, 279 (1995).

[18] see e.g. C.M. Wolfe, N. Holonyak Jr., and G.E. Stillman, in *Physical Properties of Semiconductors*, (Prentice Hall, Englewood Cliffs, New Jersey 07632, 1989) p.317

**Figure 1.** Survey spectra in normal emission of four samples of (Ga,Mn)As with varying thickness of GaAs capping from 0 to 8 MLs. The inset shows the attenuation of the Mn2p emission vs. capping layer thickness (dots) and an exponential fitting (solid line).

**Figure 2.** Mn2p spectra in normal emission of each sample prior and after As deposition with subsequent annealing.

**Figure 3.** Change of electrostatic potential at the surface of GaAs capping layer on (Ga,Mn)As as function of capping layer thickness.

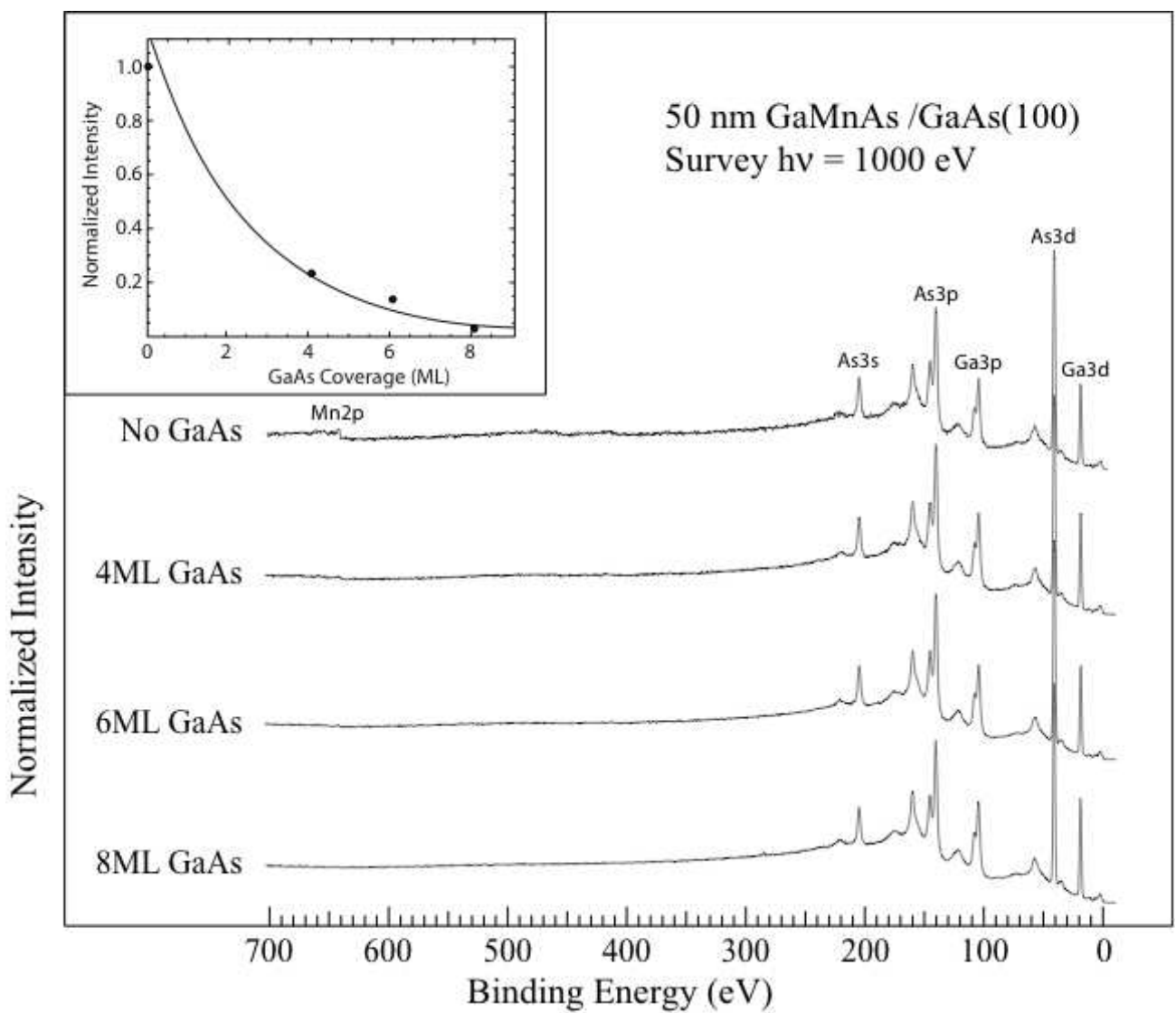

**Figure 1**

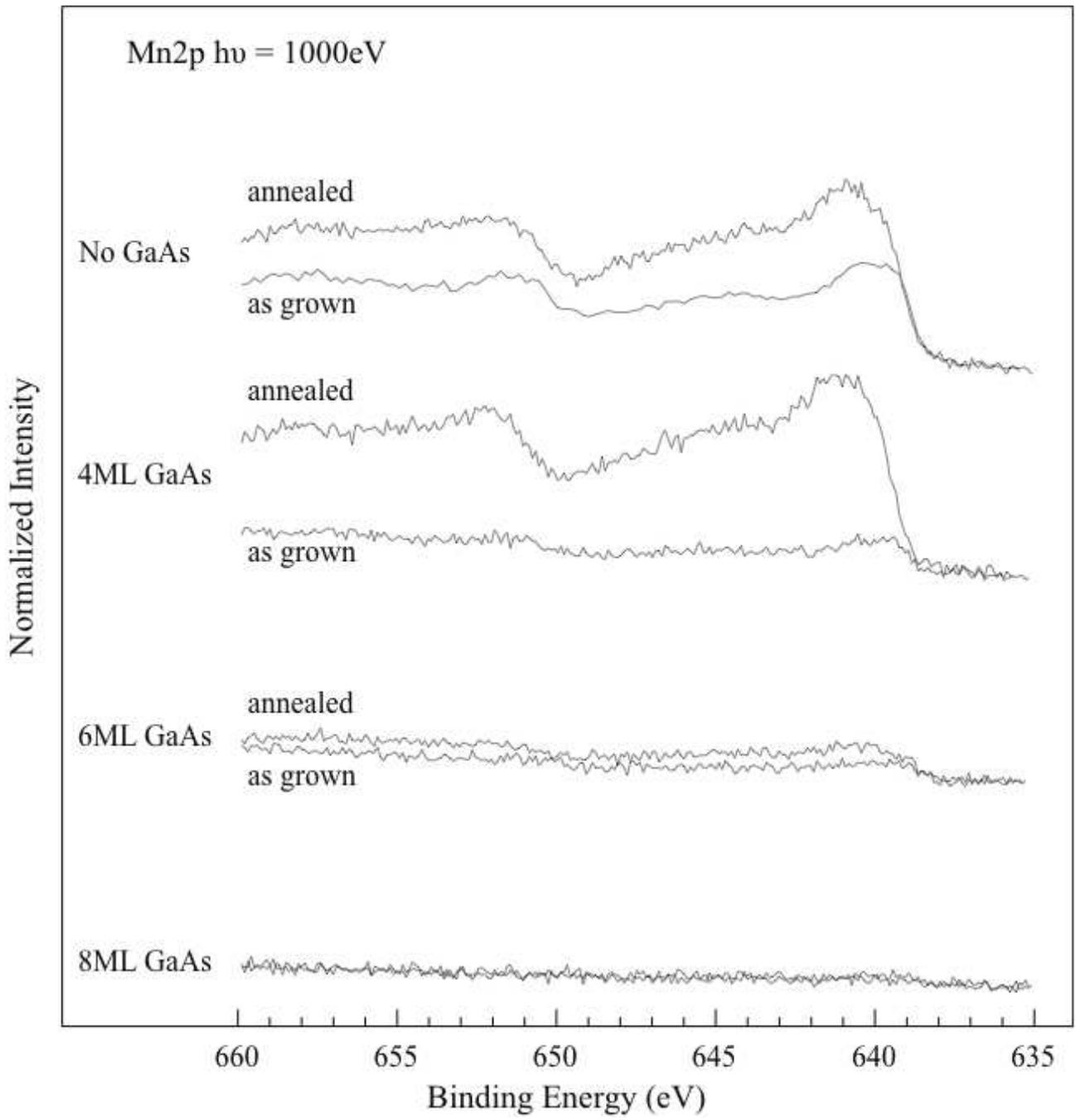

**Figure 2**

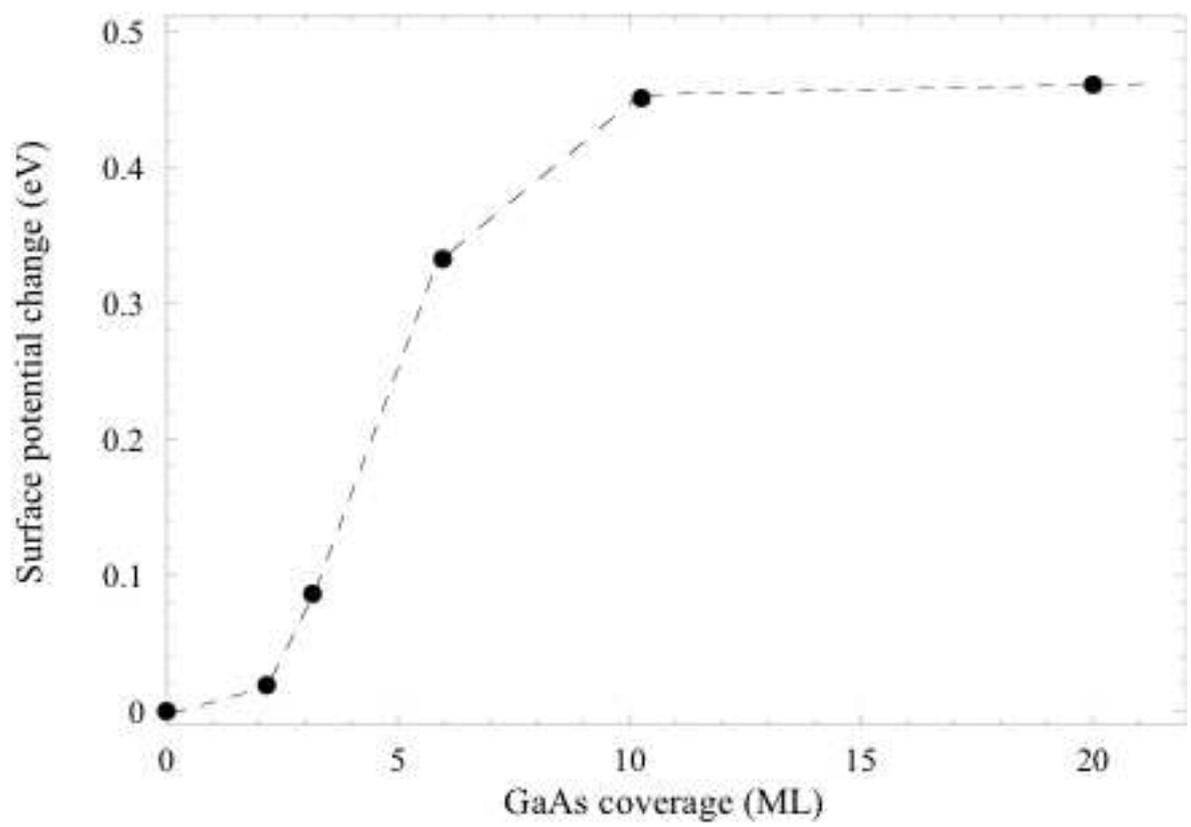

**Figure 3**